\documentclass{article}
\usepackage{amsfonts}
\usepackage{amsmath}

\input{tcilatex}
\begin{document}

\title{Path algebra algorithm for finding longest increasing subsequence}
\author{Anatoly Rodionov \\
ayarodionov@gmail.com}
\maketitle

\begin{abstract}
New algorithm for finding longest increasing subsequence is discussed. This
algorithm is based on the ideas of idempotent mathematic and uses Max-Plus
idempotent semiring. Problem of finding longest increasing subsequence is
reformulated in a matrix form and solved with linear algebra.
\end{abstract}

\section{Introduction}

The purpose of this short article is to bring attention to unifying approach
to software and hardware design suggested and developed by Grigory Litvinov,
Viktor Maslov and coworkers\cite{Litvinov}. The unifying approach is based
on observation that many algorithms do not depend on particular models of a
numerical domain and even on the domain itself. Algorithms of linear algebra
(matrix multiplication, Gauss elimination etc.) are good examples of
algorithms of this type. One can do linear algebra with field of rational
numbers, complex numbers and so on. But it is not obvious that the same
algorithms (slightly reformulated) may be used also for idempotent semirings
of different kind. Why this is interesting? Because many problems, which
never were considered as problems that have anything to do with linear
algebra, can be solved with standard linear algebra methods. For example
problem of finding shortest, critical, maximal capacity or most reliable
path on graph can be solved using linear algebra over different idempotent
semirings. Object oriented languages make it possible to reuse the same
codes for solving entirely different problems without changing a single
line! In this way all the knowledge and power of linear algebra can be used
much wider area. In fact almost all problems of dynamic programming can be
solved in this way\cite{Bellman}.

In this article I show how a very simple and classical problem of dynamic
programming -- finding of the longest increasing subsequence, can be solved
using max-plus idempotent semiring and linear algebra. The problem has
simple formulation and requires the simplest linear algebra method for its
solution. So I think that it can be used as a good illustration of the
beauty of unifying approach. Because of its simplicity the algorithm can be
already found and discussed. I never found any references on this subject.
If you know about such publication please let me know and I will add the
corresponding reference.

I used path algebra in the title to emphasize that in this article I follow
ideas of Bernard Carr\.{e} who wrote pioneer work \cite{Bernard} on using
linear algebra methods in graph theory. There many names for the same thing
now: idempotent mathematics, tropical mathematics \ldots\ Bernard Carre used 
\textit{path algebra}a \cite{Carre}. This name I prefer to use in the title
while in the article itself I will use more familiar max-plus idempotent
semiring.

\section{\protect\bigskip Formulation of the problem}

The problem is: in a given sequence of elements find length of the longest
increasing subsequence. (Here we assume that all elements of the sequence
can be compared.)

Let $S$ be a sequence of elements $S=\{s_{0},s_{1},...s_{n-1}\}$. Find the
largest length $l$ on increasing subsequence $%
\{s_{i_{0}},s_{i_{1}},...s_{i_{l-1}}\}$ such that $s_{i}\in S$ and $%
s_{i}<s_{j\text{ }}$if $i<l$.

For example for $\{5,2,8,6,3,6,9,7\}$ the longest increasing subsequence $%
\{2,3,6,9\}$ has length $4$.

The usual algorithm for solving this problem looks like:

\texttt{for i = 0, ... n-1}

\qquad L[j] = 1 + max(L[i] such that S[i] \TEXTsymbol{<} S[j])

return max(L[j])

The algorithm requires at most $O(n^{2})$ operations. (In the special case
of numerical elements algorithm of $O(n\ln (n))$ colpexity exists and even $%
O(n\ln (\ln (n))$ \cite{Bespamyatnikh}). The purpose of this note is not to
improve algorithm's speed but to show how it can be reformulated as a linear
matrix problem. For this Max-Plus algebra is used.

\section{Max-Plus algebra}

Let me recall definition of max-plus algebra.

\bigskip

Denote by $R_{\max }$ set of real number with minus infinity $R_{\max
}=R\cup \{-\infty \}$. For elements from $R_{\max }$ we can define
operations $\oplus $ and $\odot $ :

\begin{eqnarray}
a\oplus b &=&\max (a,b)  \notag \\
a\odot b &=&a+b  \notag
\end{eqnarray}

In this algebra $-\infty $ plays role of zero, and $0$\ - of unit element.
We will use symbol $\phi $ for zero and $e$ for unit element.

It can be easily proved that

\begin{itemize}
\item operation $\oplus $ is
\end{itemize}

\qquad commutative: $a\oplus b=b\oplus a$,

\qquad associative: $(a\oplus b)\oplus c=a\oplus (b\oplus c)$ ,

\qquad and idempotent: $a\oplus a=a$

\begin{itemize}
\item operation $\odot $ is
\end{itemize}

\qquad associative: $(a\odot b)\odot c=a\odot (b\odot c)$,

\qquad distributive over $\oplus $:

$\qquad a\odot (b\oplus c)=(a\odot b)\oplus (a\odot c)$ and $(b\oplus
c)\odot a=(b\odot a)\oplus (c\odot a)$.

\begin{itemize}
\item $\phi \oplus a=a$

\item $\phi \odot a=a\odot \phi =\phi $

\item $e\odot a=a\odot e=e$
\end{itemize}

\bigskip This construction is called Max-Plus algebra\footnote{%
{\footnotesize It is also called }$P_{3}${\footnotesize \ in Bernard Carr%
\.{e} classification}.} which is an example of indempotent semiring. This
algebra has many applications and was studied in many works (see for example 
\cite{Butkovic}, \cite{work}). \ 

\bigskip We will use notations $a^{n}$ for product of $n$ elements: $a^{0}=e$%
, $a^{n}=a\odot a\ldots \odot a$; $a^{+}$ for infinite sum of nonzero powers
of $a$: $a^{+}=a^{1}\oplus a^{2}\oplus a^{3}\oplus \ldots $and $a^{\ast
}=e+a^{+}$. In what follows these sums will contains only finite number of
nonzero elements.

$a^{\ast }$ gives solution for equation

\begin{equation}
y=a\odot y\oplus b
\end{equation}%
Solution is $y=a^{\ast }\odot b$. \ The result can be proved by direct
substitution and using trivial facts that $a\odot a^{\ast }=a^{+}$ and $%
a^{\ast }=e+a^{+}$. Indeed, the right hand side after the substitution is $%
a\odot a^{\ast }\odot b\oplus b=a^{+}\odot b\oplus b=(a^{+}\oplus e)\odot
b=a^{\ast }\odot b$.

\bigskip

Over elements of Max-Plus algebra we can construct matrices $A=\{a_{i,j}\}$.
We define $\oplus $ and $\odot $ operations for matrices as

\begin{eqnarray*}
A\oplus B &=&\{a_{i,j}\oplus b_{i,j}\} \\
A\odot B &=&\left\{ \oplus \sum_{k=0}^{n}a_{i,k}\odot b_{k,j}\right\}
\end{eqnarray*}%
where $\oplus \sum_{k=0}^{n}a_{k}=a_{0}\oplus a_{1}\oplus \ldots a_{n}$.
Zero element $\Phi $ is matrix filled with $\phi $; unit element $E$ -
matrix with $e$ on diagonal and $\phi $ in all other positions.

As in scalar case define $A^{n}$ for product of $n$ elements: $A^{0}=E$, $%
A^{n}=A\odot A\ldots \odot A$; $A^{+}$ for infinite sum of nonzero powers of 
$A$: $A^{+}=A^{1}\oplus A^{2}\oplus A^{3}\oplus \ldots $and $A^{\ast
}=E+A^{+}$.

The constructed algebra of matrices is itself an idempotent semiring.

In this algebra equation

\begin{equation}
Y=A\odot Y\oplus B  \label{MATREQ}
\end{equation}%
has solution $Y=A^{\ast }\odot B$.

If $B=E$ then $Y=A^{\ast }$. This fact will be used later in the algorithm
for efficient calculation of $A^{\ast }$.

\section{Path algebra algorithm}

The algorithm contains two steps: constructing of incidence matrix and
solving linear equation.

\subsection{Constructing of incidence matrix}

\bigskip We will construct graph which vertices represents elements of the
sequence $S$. Two vertices $s_{i}$ and $s_{j\text{ }}$are connected by
directed edge $s_{i}\rightarrow s_{j\text{ }}$if $s_{i}<s_{j}$ and $i<j$.

We will use incidence matrix for graph representation. Element $a_{i,j}$ of
incidence matrix is equal to $1$ if $s_{i}$ and $s_{j\text{ }}$are
connected, else $\phi $:

\begin{equation*}
a_{i,j}=1\text{ if }s_{i}<s_{j}\text{ and }i<j\text{ else }\phi
\end{equation*}

By construction this matrix is upper triangle.

For the example above $\{5,2,8,6,3,6,9,7\}$ the matrix looks like:

\begin{equation*}
A=%
\begin{bmatrix}
\phi & \phi & 1 & 1 & \phi & 1 & 1 & 1 \\ 
\phi & \phi & 1 & 1 & 1 & 1 & 1 & 1 \\ 
\phi & \phi & \phi & \phi & \phi & \phi & 1 & \phi \\ 
\phi & \phi & \phi & \phi & \phi & \phi & 1 & 1 \\ 
\phi & \phi & \phi & \phi & \phi & 1 & 1 & 1 \\ 
\phi & \phi & \phi & \phi & \phi & \phi & 1 & 1 \\ 
\phi & \phi & \phi & \phi & \phi & \phi & \phi & \phi \\ 
\phi & \phi & \phi & \phi & \phi & \phi & \phi & \phi%
\end{bmatrix}%
\end{equation*}

Matrix \ $A$ describes all path of length one in the graph, matrix $A^{2}$

\begin{equation*}
A^{2}=%
\begin{bmatrix}
\phi  & \phi  & \phi  & \phi  & \phi  & 2 & 2 & 2 \\ 
\phi  & \phi  & \phi  & \phi  & \phi  & \phi  & \phi  & \phi  \\ 
\phi  & \phi  & \phi  & \phi  & \phi  & \phi  & \phi  & \phi  \\ 
\phi  & \phi  & \phi  & \phi  & \phi  & \phi  & 2 & 2 \\ 
\phi  & \phi  & \phi  & \phi  & \phi  & \phi  & \phi  & \phi  \\ 
\phi  & \phi  & \phi  & \phi  & \phi  & \phi  & \phi  & \phi  \\ 
\phi  & \phi  & \phi  & \phi  & \phi  & \phi  & \phi  & \phi  \\ 
\phi  & \phi  & \phi  & \phi  & \phi  & \phi  & \phi  & \phi 
\end{bmatrix}%
\end{equation*}%
describes all paths of length two and so on. Because the graph is acyclic
all paths are shorter that number of vertices $N$, so matrix $A^{N}$
contains only zero elements.

Indeed, the problem may be solved by subsequently calculating powers of
matrix $A$: $A^{2}$, $A^{i}$, \ldots ; the last power $k$ such that $A^{i}$
contains at least one element not equal to $\phi $ is the required maximum
length minus one. This solves the problem. For the example above the power
is equal to $3$.

Unfortunately the cost of this solution is $N^{3}$: in the worst case we
have to make $N$ matrix multiplications and each multiplication requires $%
N^{2}$ operations. There is a more efficient way to find the maximum length
- solving linear equation (\ref{MATREQ}).

\subsection{Solving linear equation}

The $n$-th power of matrix $A$ contains all paths of length $n$, so sum of
powers of matrix $A$ (which is $A^{\ast }$) contains all possible maximum
length paths in graph. We can add zero power of $A$ which by definition
equals to unity matrix $E$. Unity matrix has Max-Plus unit elements on its
diagonal which are real zeros, so we may think about $A^{0}$ as matrix of
paths of zero length connecting edges to themselves. Adding $A^{0}$ to $A^{+}
$ gives $A^{\ast }$. So $A^{\ast }$ contains all maximum length paths. If we
know $A^{\ast }$ then can find length of the longest path. But direct
calculation of $A^{\ast }$ is costly operation.

To solve the problem let's use (\ref{MATREQ}) with $B=E$.

\begin{equation*}
Y=A\odot Y\oplus E
\end{equation*}%
Solution of this equation is  $Y=A^{\ast }$.

Let $Y^{i}$ be the $i$-th column of matrix $Y$ and $E^{i}$ be the $i$-th
column of matrix $E$. Then the equation can be rewritten as system of $N$
equations:

\begin{equation*}
Y^{i}=A\odot Y^{i}\oplus E^{i}\text{ where }i=0,\ldots ,N
\end{equation*}

Elements $y_{i}^{j}\in Y^{j}$ are maximum path length from $s_{i}$ to $s_{j} 
$. Elements $x_{j}$ of vector $X=$ $Y^{0}\oplus Y^{1}\oplus \ldots Y^{N-1}$
(here superscript is index, not power) give maximum length to element $s_{j} 
$. Equation to $X$ is:

\begin{equation}
X=A\odot X\oplus U
\end{equation}%
where $U$ is $N$ column and one row matrix each all elements are equal to $e$%
.

Because $A$ is upper diagonal matrix the equation can be easily solved by
direct substitutions:

\begin{eqnarray*}
L &=&\oplus \sum_{i=0}^{N-1}x_{i}\text{ where} \\
x_{N-1} &=&e \\
x_{i} &=&e\oplus (\oplus \sum_{j=i+1}^{N-1}a_{i,j}\odot x_{j})\text{ where }%
(i=N-2,\ldots ,0)
\end{eqnarray*}%
(here $\oplus \sum $ is sum over operation $\oplus $). $L+1$ gives maximum
length on an increasing subsequence in $S$.

This formula gives us solution in $\frac{N(N+1)}{2}$ time which is the same
as in classical dynamic programming algorithm. In fact the last formula can
be rewritten in the usual way:

\begin{eqnarray*}
L &=&\max_{i=0,\ldots ,N-1}\left( x_{i}\right) \text{ where} \\
x_{N-1} &=&0 \\
x_{i} &=&\max_{j=i+1,\ldots ,N-1}(a_{i,j}+x_{j})\text{ where }(i=N-2,\ldots
,0)\text{ and }a_{i,j}\neq 0
\end{eqnarray*}

\section{New class of generic algorithms}

\bigskip\ As it was pointed out by David Musser and Alexander Stepanov 
\textit{"generic programming centers around the idea of abstracting from
concrete, efficient algorithms to obtain generic algorithms that can be
combined with different data representations to produce a wide variety of
useful software" }\cite{Stepanov}.

But what are these generic algorithms? How to find them? The great success
of standard template library (STL)  is based greatly on containers and
algorithms specific for containers - iteration through container, finding
elements, sorting. But what are other classes of generic algorithms? Paths
algebra gives an answer to this question - linear algebra algorithms used on
idempotent semirings. The fact that all dynamic programming algorithms can
be reformulated in terms of linear algebra should not be overlooked. In
fact, we already have many different parametrized representation of vectors
and matrices (for example in BOOST \textit{C++} library\cite{BOOST}). But
what kind of elements for these matrices can we picked in BOOST? Integer
numbers, floating point numbers, rational numbers, complex numbers,
octonions, quaternions, intervals... - all from so called numerical domain.
Why don't we add some more interesting objects - semirings? As it was shown
in \cite{Litvinov} and \cite{Bellman} this can be very useful for solving a
wide variety of problems.

\end{document}